\documentclass[runningheads]{llncs}
\usepackage[T1]{fontenc}
\usepackage{graphicx}
\usepackage{amssymb}
\usepackage{amsmath}

\usepackage{xcolor}

\usepackage{tikz}
\usetikzlibrary{arrows.meta,positioning,fit,decorations.pathreplacing}

\usepackage{braket}

\usepackage{booktabs}
\usepackage{tabularx}
\usepackage{array}

\usepackage{hyperref}
\hypersetup{hidelinks}

\begin{document}
\title{Practical Considerations for Processing Massive Classical Data via Quantum Oracle Sketching}
%
\titlerunning{Practical Considerations for Processing Massive Classical Data via QOS}
%
\author{Stefano Markidis  } 
\authorrunning{S. Markidis}
%
\institute{KTH Royal Institute of Technology, Stockholm, Sweden 
}
\maketitle              
\begin{abstract}
Classical-to-quantum I/O is a critical bottleneck for data-intensive quantum workloads. Quantum Oracle Sketching (QOS) has been proposed to address this
problem. We identify its practical requirements, including the representation of data as a probability distribution over oracle addresses, the relation
between sample count, approximation error, and update cost, and the coherence and runtime support required for online execution. To study QOS on current
quantum computing devices, we develop an offline realization that compiles sample blocks into empirical phase oracles and evaluate it using DNA fingerprinting.
The application converges with relatively few samples, while exact phase fusion combines repeated updates to the same address. However, the compiled circuits
remain too deep for current quantum computing devices. Practical QOS therefore requires lower depth oracle synthesis and interactive quantum runtimes.
\end{abstract}

\section{Introduction}
Classical-to-quantum I/O is a critical bottleneck for data-driven quantum algorithms, particularly quantum machine learning. Classical features, observations, or probability distributions must first be made accessible to the quantum computation through state preparation, a circuit, or an oracle~\cite{Biamonte2017QML,MarinSanchez2023FunctionLoading}. Many algorithmic analyses assume that this interface has already been constructed and exclude its cost from the reported complexity. However, for large classical inputs,  constructing the interface may dominate the end-to-end computation and reduce or eliminate the expected quantum advantage~\cite{Aaronson2015ReadFinePrint}. This is the quantum analogue of the classical I/O bottleneck. Classical streaming and sketching algorithms address the corresponding problem by processing inputs sequentially and retaining compact summaries for selected queries~\cite{Alon1996AMS}.

Related quantum methods also seek compact representations of classical data. Quantum fingerprinting uses short quantum states to compare inputs~\cite{Buhrman2001Fingerprinting}, while quantum sketching extends this idea to the estimation of similarity and distance measures~\cite{Doriguello2019QuantumSketching}. These methods reduce the stored representation, but the corresponding quantum states must still be constructed from the classical input. Other approaches assume coherent data access interfaces. QRAM architectures provide queries to classically stored data in quantum superposition~\cite{Giovannetti2008QRAM,Giovannetti2008Arch}, although their implementation, noise sensitivity, and fault-tolerant resource requirements remain important challenges~\cite{Arunachalam2015QRAM,Hann2021QRAM}. QROM replaces active memory access with a circuit representation of fixed classical data~\cite{Babbush2018QROM}, while compact QRAM constructions can support related block-encoding operations~\cite{Duan2024CompactQRAM}. Constructing any of these interfaces remains a major quantum-classical systems challenge.

Quantum Oracle Sketching (QOS) was recently proposed as an alternative model for constructing quantum data access from classical samples~\cite{Zhao2026QOS}. For selected applications and data models, the QOS analysis shows that the resulting quantum algorithms operate with exponentially less working memory than the classical machines covered by the corresponding lower bounds. When the data generating distribution changes over time, the analysis further shows that classical machines operating under the stated memory constraints
require superpolynomially more samples and computation time to match the quantum result. These separations apply to an online model in which samples are processed and discarded during quantum execution. Therefore, their realization depends on both the quantum implementation of the sampled oracle and the interface between the classical data stream and the active quantum computation.

This paper examines the practical requirements of QOS from a computer science perspective. We organize the analysis around the representation of classical data as a probability distribution, the sample count required to approximate the oracle, repeated oracle applications, the choice and implementation of the oracle family, and the coherence and runtime requirements of online execution. We then develop an offline formulation compatible with the batched circuit execution model used by current quantum platforms. A DNA sequence fingerprinting use case evaluates the accuracy and circuit cost of the offline formulation. 

\section{Quantum Oracle Sketching}
\begin{figure}[t]
    \centering
    \includegraphics[width=\textwidth]{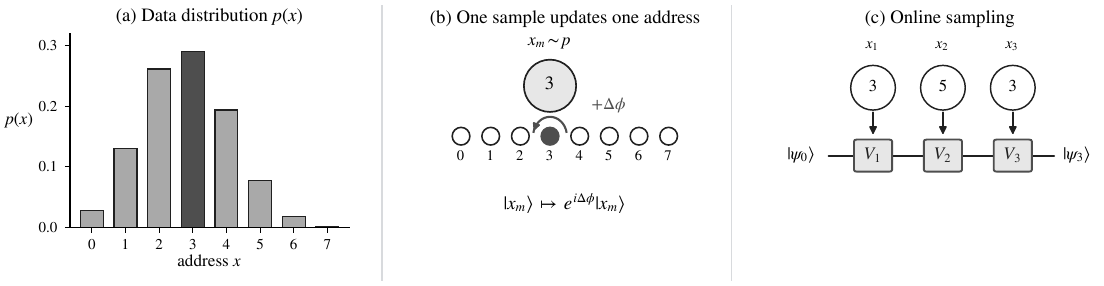}
    \caption{Three elements of online QOS.
    (a) A data distribution \(p(x)\), where each bin or category defines
    an oracle address.
    (b) A sample \(x_m\sim p\) selects the address receiving a local
    phase update.
    (c) Samples are processed sequentially and discarded after their
    updates, while the quantum state is retained.}
    \label{fig:qos_principles}
\end{figure}

QOS separates the oracle required by a quantum algorithm from its implementation. The algorithm is formulated using an ideal oracle \(U\), which QOS approximates through quantum updates controlled by classical samples. Fig.~\ref{fig:qos_principles} provides an overview of the phase construction. A data distribution determines which addresses are sampled, and each sample controls a local phase update. In the online model, samples are processed during quantum execution and discarded after use.

The first step is to define what the oracle addresses represent. For numerical data, an address may identify a histogram bin. For categorical data, it may identify a category or class. For a finite data set, \(p(x)\) is the fraction of observations assigned to address \(x\).
Each bar in Fig.~\ref{fig:qos_principles}(a) corresponds to one address, and its height gives this probability. The quantum register represents
address \(x\) by the computational basis state \(\ket{x}\). An address may also carry a real valued weight \(f(x)\). In the simplest
case, \(f(x)=1\), and the oracle represents only the probabilities
\(p(x)\). For simplicity, we first explain QOS using a diagonal phase oracle.
QOS also supports other constructions, including state preparation
and block encodings, which we discuss later in this section. The phase construction assigns the angle \(t p(x)f(x)\) to
each address, where \(t\) sets the overall phase scale:
\begin{equation}
U(t)
=
\sum_x
e^{\mathrm{i}t p(x)f(x)}
\ket{x}\bra{x}.
\label{eq:qos_ideal_phase_oracle}
\end{equation}
This is a diagonal phase oracle. Each address receives its own
phase factor, with no transfer of amplitude between addresses. The data are therefore encoded in relative phases.

QOS approximates this oracle by sampling addresses from \(p(x)\). For a finite data set, this can be done by selecting an observation uniformly at random and returning its assigned address. Explicitly constructing and storing the full histogram is not necessary. Each sample produces a small phase update at the selected address,
as illustrated in Fig.~\ref{fig:qos_principles}(b). Repeated updates accumulate to approximate the phase angles of the ideal oracle. In online execution~\cite{Zhao2026QOS}, each update is applied as its sample arrives. The sample is then discarded, while its contribution remains in the evolving quantum state, as shown in Fig.~\ref{fig:qos_principles}(c). \\[0.4em]
\noindent
\textbf{Practical Consideration I: QOS encodes data as a probability
distribution, not item by item.}
\textit{The application must define what each address represents, how
addresses are sampled, and any associated weights \(f(x)\). Preparing this representation may require binning, categorization, aggregation, or reweighting. These choices determine which information from the original data remains available to the query algorithm. Any preprocessing costs, together with the cost of generating samples,
must be included in the complete QOS data access pipeline.}
\par\vspace{0.8em}

Once the data distribution has been specified, the next choice is how many samples to use for one oracle application. Let \(M_0\) denote this sample count. QOS draws \(M_0\) independent addresses from \(p(x)\) and applies one phase update per sample:
\begin{equation*}
V_m
=
\exp\left(
\mathrm{i}\frac{t f(x_m)}{M_0}
\ket{x_m}\bra{x_m}
\right),
\qquad
V=V_{M_0}\cdots V_1.
\end{equation*}
Each update changes the phase only at the sampled address. The factor \(1/M_0\) keeps the target phase scale fixed. Increasing
the sample count produces more, smaller updates rather than a larger overall phase.

After all updates, the accumulated phase at address \(x\) equals \(t f(x)\) multiplied by the fraction of samples that selected that
address. This fraction estimates \(p(x)\), so the accumulated phase approximates the ideal angle \(t p(x)f(x)\). An address that is not
sampled receives no update, while repeated samples at the same address contribute additional phase increments.

For a finite sample count, the estimated probabilities fluctuate, and different sample blocks produce different approximate oracles.
The QOS guarantee bounds the error of the quantum channel averaged over independently sampled blocks~\cite{Zhao2026QOS}. Increasing \(M_0\) tightens this bound, but also increases the number of samples processed and quantum updates applied.

The required sample count need not equal the number of observations
in the original data set. QOS estimates the probabilities defining the
oracle rather than encoding every observation individually. For a
very large data set, these probabilities may be approximated to the
required accuracy using far fewer samples. The achievable reduction
depends on the distribution, the phase scale and weights, and the
accuracy needed by the downstream query.\\[0.4em]
\noindent
\textbf{Practical Consideration II: Sampling trades oracle accuracy for fewer data encoding updates.}
\textit{QOS can approximate the distribution of a large data set using a much smaller sequence of phase updates. The practical question is how short this sequence can be while preserving the accuracy of the downstream query. Reducing the sample count reduces the number of updates but increases uncertainty in the encoded phases. Therefore, the useful operating point depends on how sensitive the query is to these errors.}
\par\vspace{0.8em}

The sample count determines how many updates are applied. In online QOS, samples are consumed one at a time during quantum execution, as illustrated in
Fig.~\ref{fig:qos_principles}(c). The current sample supplies the address
and value needed for its phase update. Once the update has been applied,
the sample can be discarded and the same working registers reused for
the next sample.

What persists is the evolving quantum state, which carries the effect of the updates already applied. The processor does not need to retain the sample history or a list of all operations. Therefore, increasing the sample count lengthens the update sequence without requiring more sample records to be held simultaneously. Any storage required by the original data source or its sampling procedure remains a separate part of the complete system cost.

This distinction between retained memory and processing work is central to QOS. Constructing one oracle application still requires processing \(M_0\) samples and applying their updates. The machine size separations proved for selected QOS applications rely on this online resource model~\cite{Zhao2026QOS}, in which samples affect the quantum computation without accumulating in the processor's memory.\\[0.4em]
\noindent
\textbf{Practical Consideration III: The memory advantage relies on
applying samples online and discarding them after use.}
\textit{The key saving is that a long sample stream need not become a large stored data structure or a precompiled sequence of sample dependent operations. Each sample is used once, while the quantum state is retained across updates. Collecting samples or materializing the complete oracle before execution changes this memory model, so the online machine size guarantees do not directly carry over.}
\par\vspace{0.8em}

The construction described so far implements one oracle application. To use QOS within a complete algorithm, we must identify where the algorithm accesses the data and what quantum processing occurs between those accesses. A useful example is \emph{data re-uploading} \cite{perez2020data}, where data encoding operations alternate with fixed or trainable quantum transformations. The transformations mix the quantum state, so each new encoding acts on the result of the preceding computation. 

For \(Q\) oracle applications, the ideal algorithm and its sampled realization can be written as
\begin{equation}
\begin{aligned}
A  &= C_Q U_Q \cdots C_1 U_1 C_0,\\
A' &= C_Q V_Q \cdots C_1 V_1 C_0.
\end{aligned}
\label{eq:qos_query_composition}
\end{equation}
Here, \(U_q\) is an ideal data oracle, \(V_q\) is its sampled
approximation, and \(C_q\) denotes the surrounding quantum processing.
QOS replaces the oracle applications while leaving these intervening
operations unchanged. If the \(C_q\) contain trainable parameters,
those parameters are fixed for a given execution and independent of
the sample blocks used to construct the oracles.

Even when several applications use the same ideal oracle, the
independent block QOS construction assigns a fresh sample block
\(B_q\) to each application. With \(M_0\) samples per block, the complete
execution consumes \(M=QM_0\) samples. Fresh blocks mean independent
draws, not necessarily distinct observations or addresses.
Reusing a single sampled oracle across applications would correlate
their approximation errors and requires a separate error analysis.

\begin{figure*}[t]
    \centering
    \includegraphics[width=\textwidth]{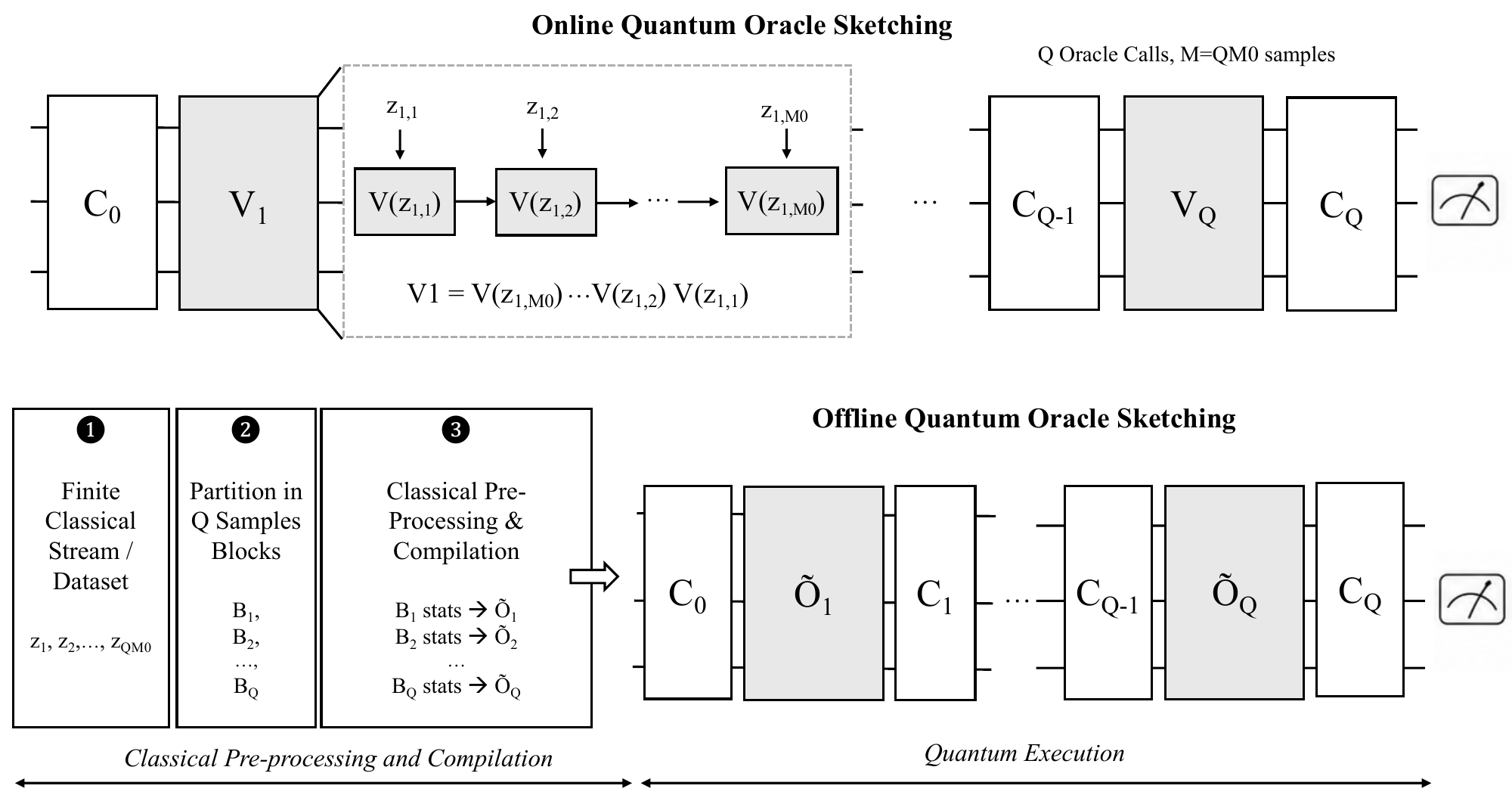}
    \caption{Online and offline QOS with \(Q\) oracle occurrences and
    \(M=QM_0\) samples. Each \(U_q\) is replaced using an independent
    block \(B_q\): online QOS constructs \(V_q\) during execution,
    whereas offline QOS compiles \(\widetilde O_q\) beforehand.}
    \label{fig:qos_streaming_vs_offline}
\end{figure*}

Fig.~\ref{fig:qos_streaming_vs_offline} compares the online and
offline realizations of this sequence. In the top panel, each
\(V_q\) is applied through a sequence of updates
during quantum execution. The expanded first oracle shows these
individual updates. After completing an oracle application, execution
continues through the intervening quantum operations before consuming
the next sample block. The lower panel shows the offline construction
developed in the next section.\\[0.4em]
\noindent
\textbf{Practical Consideration IV: Data access must be isolated in explicit oracle calls.}
\textit{Applying QOS requires identifying which operations introduce the data and expressing them as compatible quantum oracles. For a data re-uploading circuit, these are the encoding operations between the fixed or trainable transformations. Each encoding occurrence receives its own sampled realization under the independent block construction.}
\par\vspace{0.8em}

Once the data dependent operations have been identified, the next question is what each oracle must do. Encoding data in phases, preparing a quantum state, and providing access to matrix entries are different operations. An algorithm designed for one of these interfaces cannot generally use another without modification.

In the phase construction considered here, each sample supplies
an address and its associated value \(f(x)\). These determine where
the update acts and how much phase it adds. This gives a direct
connection between a classical sample and a quantum operation.
Because the updates are diagonal in the same address basis, they
commute. Repeated updates to an address can therefore be combined
by summing their phase angles within one oracle application.

The original QOS formulation also considers state preparation,
sparse matrix access, and block encodings~\cite{Zhao2026QOS}.
State preparation constructs a normalized quantum state whose
amplitudes represent the data, for example using the uniformly
controlled rotations of M{\"o}tt{\"o}nen
et al.~\cite{Mottonen2005StatePreparation}. A block encoding instead embeds a scaled matrix within a larger unitary, providing an interface for quantum matrix algorithms~\cite{Gilyen2019QSVT}. For each construction, the sampling procedure must supply the information required by that interface.

Even for a phase oracle, a simple sample update need not correspond
to a simple hardware operation. An update is local to one address, not necessarily to one qubit. Selecting that address can require controls across the quantum register, so a single update may expand into many elementary gates. \\[0.4em]
\noindent
\textbf{Practical Consideration V: The oracle family determines
the sample requirements and circuit structure.}
\textit{The oracle must provide the transformation expected by the algorithm. Its choice determines what information each sample must carry and which normalization, auxiliary registers, and quantum operations are needed. The simple a phase update, used in this work, applies to the diagonal phase construction. Other oracle families require their own sampled implementations. Therefore, resource estimates must be made for the chosen oracle.}
\par\vspace{0.8em}

Online QOS imposes two distinct requirements: the quantum state must survive the complete streaming computation, and the execution interface must allow incoming samples to control that computation.

The quantum state must remain coherent throughout the complete streaming computation. This includes the time required for quantum operations, sample
delivery, and real-time classical control. Without error correction, the total duration must be much shorter than the effective coherence time of the quantum state. A long stream can violate this condition even when each individual update is fast. Feasibility therefore requires device-specific estimates of gate durations, sample delivery rates, and control latency. Fault tolerance replaces this physical coherence constraint with the requirement to maintain sufficiently low logical error throughout the computation.

Online QOS requires interaction within a single quantum execution. New classical samples must determine subsequent operations while the existing query state is retained. This differs from a workflow in which circuits and their inputs are submitted as jobs and results are returned after execution. Updating parameters between jobs, or grouping jobs into a batch or session, does not by itself provide interactive execution. OpenQASM~3 represents progress toward this model through classical control flow, external classical function calls, and explicit timing~\cite{Cross2022OpenQASM3}. However, language support does not guarantee that a particular backend can accept external samples at the required rate. \\[0.4em]
\noindent
\textbf{Practical Consideration VI: Online QOS requires sufficient coherence and an interactive execution interface.}
\textit{First, the quantum state must remain sufficiently accurate throughout all updates, quantum gates, and waits for classical samples. Second, the controller must allow those samples to influence the same ongoing quantum execution. Neither a short gate sequence alone nor support for dynamic control flow establishes both requirements. }

\section{Offline Quantum Oracle Sketching}
The requirements of the previous practical consideration motivate the offline realization developed next. We  move sample processing to the classical host and construct the sampled oracles before quantum execution. The lower panel of Fig.~\ref{fig:qos_streaming_vs_offline} shows the resulting
separation between classical preprocessing and quantum execution.

The sampling procedure remains unchanged. Each oracle application receives an independent block \(B_q\) containing \(M_0\) samples. For each block, the host accumulates the phase contributions at the sampled addresses and compiles the resulting summary into an empirical oracle circuit \(\widetilde O_q\). Samples can be processed sequentially and discarded after updating the summary;
the raw sample block need not be retained.

The compiled oracles replace the corresponding data access operations in Eq.~\eqref{eq:qos_query_composition}, giving
\begin{equation}
A_{\mathrm{off}}
=
C_Q \widetilde O_Q \cdots C_1 \widetilde O_1 C_0.
\end{equation}
This preprocessing and compilation occur entirely on the classical host. During quantum execution, the data-dependent gates and their parameters are already specified, so no further samples need to arrive from the classical source.

The compilation target is the sampled oracle \(V_q\), not the ideal oracle \(U_q\). With exact synthesis, \(\widetilde O_q=V_q\), and the offline circuit implements the same
transformation as the corresponding online sample sequence.
Consequently, the sampling approximation and the independence
between oracle applications are preserved. Any approximation
introduced by circuit synthesis is an additional error, separate
from sampling error. Hardware noise contributes a further source
of error during execution.

Therefore, the change is in how the sampled transformation is
represented and delivered.  Offline QOS  retains a
classical summary during construction and materializes a circuit
description before execution. Although both approaches can discard
raw samples, they do not have the same memory requirements.
Offline feasibility depends on the storage and processing needed
to build the circuits, as well as their final gate counts and
depths. \\[0.4em]
\noindent
\textbf{Practical Consideration VII: Offline QOS preserves the
sampled construction, not the online memory model.}
\textit{With exact synthesis, the compiled circuit reproduces the quantum updates and retains their sampling error guarantee. However, the circuit descriptions introduce storage and compilation costs absent from the original online construction. Therefore, the online machine size separations do not directly apply. An offline implementation must be evaluated
as a complete classical-quantum computation, including oracle construction, program storage, and quantum execution.}
\par\vspace{0.8em}

Offline compilation can simplify a sampled oracle before gate synthesis when its updates can be combined exactly. For the diagonal phase oracle considered here, the updates commute. Repeated updates to the same address can therefore be replaced by one operation carrying their accumulated phase. We call this exact compilation optimization \emph{phase fusion}. It changes neither the empirical oracle nor its sampling error, and the number of samples processed remains unchanged. Other oracle families require a separate analysis of which operations, if any, can be combined.

The effectiveness of fusion depends on the sampling distribution. A peaked distribution repeatedly selects a small set of addresses and
therefore provides more opportunities for fusion. A flatter distribution spreads the samples across more addresses and generally provides less
compression at the same sample count. Repetition can nevertheless remain substantial when the sample count is large relative to the address space.

Fusion is performed separately within each oracle application. The transformations \(C_q\) between applications generally do not commute with the phase oracle, so combining updates across them would change the computation. The reduction in phase operations is therefore a logical circuit optimization; its effect on native gate counts and depth must be evaluated after synthesis and transpilation. \\[0.4em]
\noindent
\textbf{Practical Consideration VIII: Fusion depends on the oracle structure and the sampling distribution.}
\textit{Phase fusion is an exact compilation optimization. Its benefit depends on whether the quantum operations can be combined and on how frequently the same addresses occur within each sample block. Peaked distributions generally provide stronger fusion than flatter distributions, while quantum gate transformations prevent fusion across oracle applications. .}
\par\vspace{0.8em}

\section{Case Study: DNA Sequence Fingerprinting}
\label{sec:dna_use_case}
We use DNA sequence retrieval to test whether sampled oracles preserve useful similarity information while reducing the quantum encoding operations. For each query sequence, the task is to retrieve another sequence derived from the same parent. A synthetic corpus provides known family labels, allowing us to compare retrieval using exact frequency data with retrieval using offline QOS.

A DNA sequence is represented by the frequencies of its \(k\)-mers, the contiguous substrings of length \(k\). These form a categorical data distribution. Each possible \(k\)-mer defines a category, and its relative frequency gives the corresponding probability. This representation retains local substring composition but discards substring positions. The use case therefore concerns composition-based retrieval.

For \(N\) sequences of length \(L\), there are \(d=4^k\) possible \(k\)-mers over \(\{A,C,G,T\}\). We write \(X_{ij}\) for the fraction of the \(L-k+1\) overlapping windows in sequence \(i\) that contain \(k\)-mer \(j\). Therefore, each row of the frequency matrix \(X\) sums to one. An oracle address is the pair \((i,j)\), identifying both a sequence and a \(k\)-mer category.

We construct phase fingerprints by assigning an angle \(\alpha X_{ij}\) to each address. For a fixed sequence \(i\), the feature-register oracle \(O_{X,i}\) acts on an equal superposition \(\ket{+}_d\) of the \(k\)-mer categories:
\begin{equation}
\begin{aligned}
O_X\ket{i,j}
&=
e^{\mathrm{i}\alpha X_{ij}}\ket{i,j},\\
\ket{\phi_i}
&=
O_{X,i}\ket{+}_d
=
\frac{1}{\sqrt d}
\sum_{j=0}^{d-1}
e^{\mathrm{i}\alpha X_{ij}}\ket{j}.
\end{aligned}
\label{eq:dna_phase_fingerprint}
\end{equation}
Similarity is measured by the fidelity, or squared overlap, between fingerprint states. The scale \(\alpha\) controls sensitivity to frequency differences. However, the periodic encoding does not guarantee a globally monotone relation between fidelity and frequency-vector distance. We evaluate the fidelities directly, without a small-phase approximation.

To include repeated data access, we use a data re-uploading pattern for benchmarking purposes. Phase encoding alternates with fixed Hadamard transformations. Writing \(H_d=H^{\otimes\log_2 d}\), the reference state after
\(Q\) encoding applications and its pairwise fidelity are
\begin{equation}
\begin{aligned}
\ket{\psi_i^{(Q)}}
&=
\left(H_dO_{X,i}\right)^Q\ket{+}_d,\\
F_{\mathrm{exact}}^{(Q)}(i,\ell)
&=
\left|
\langle\psi_i^{(Q)}|\psi_\ell^{(Q)}\rangle
\right|^2.
\end{aligned}
\label{eq:dna_exact_q_query_state}
\end{equation}
For one application, the final Hadamard preserves inner products, giving the same fidelities as the phase fingerprints in Eq.~\eqref{eq:dna_phase_fingerprint}. With repeated applications, the intermediate Hadamards mix feature amplitudes and generally prevent the encodings from being combined into one phase oracle.

The QOS sampler selects a sequence uniformly and then selects one of its \(k\)-mer windows uniformly. It returns the address \((i,j)\), where \(j\) is the observed \(k\)-mer. The resulting joint probability is \(p(i,j)=X_{ij}/N\). Samples can therefore be drawn directly from the sequences without first constructing the complete frequency matrix.

For oracle application \(q\), the host counts the sampled addresses in an independent block of \(M_0\) samples. If \(C_{q,ij}\) is the count at address \((i,j)\), multiplying its observed joint frequency by \(N\) gives an unbiased estimate of \(X_{ij}\). The empirical oracle and sampled fingerprint state are
\begin{equation}
\begin{aligned}
\widehat X_{q,ij}
&=
\frac{NC_{q,ij}}{M_0},\\
V_q\ket{i,j}
&=
e^{\mathrm{i}\alpha\widehat X_{q,ij}}\ket{i,j},\\
\ket{\widetilde\psi_i^{(Q)}}
&=
H_dV_{Q,i}\cdots H_dV_{1,i}\ket{+}_d.
\end{aligned}
\label{eq:dna_sampled_q_query_state}
\end{equation}
Each sample contributes a phase \(\alpha N/M_0\) at its address. This is the QOS phase construction with unit weights and \(t=N\alpha\). Fresh independent blocks are used for successive applications, giving a total sample count \(M=QM_0\). Offline phase fusion combines repeated addresses within each block while preserving the intervening Hadamard transformations.

The data set contains \(N=32\) sequences: eight mutated variants from each of four independently generated parents. The sequence length is \(L=512\), and the GC-content parameter is \(0.5\). Each base is replaced independently with probability \(0.05\), choosing uniformly among the other three bases. We use \(k=4\), giving \(d=256\) categories, and set \(\alpha=8\). 

We evaluate \(Q\in\{1,2,4,8\}\) and \(M\in\{2^{10},2^{11},2^{12},2^{13},2^{14},2^{15}\}\), with \(M_0=M/Q\). For each parameter pair, we generate 32 independent sampled constructions. Each is compared with the exact reference using the same number of encoding applications. Accuracy is computed from ideal state vectors with exact empirical phase operations. This isolates finite sample error from measurement noise, hardware noise, and synthesis approximations.

We measure both agreement with the exact fingerprint matrix and the ability to retrieve the correct family. The sampled matrix and its relative Frobenius error are
\begin{equation}
\begin{aligned}
F_{\mathrm{QOS}}^{(Q)}(i,\ell)
&=
\left|
\langle
\widetilde\psi_i^{(Q)}
|
\widetilde\psi_\ell^{(Q)}
\rangle
\right|^2,\\
E_F^{(Q)}
&=
\frac{
\left\|
F_{\mathrm{QOS}}^{(Q)}
-
F_{\mathrm{exact}}^{(Q)}
\right\|_F
}{
\left\|
F_{\mathrm{exact}}^{(Q)}
\right\|_F
}.
\end{aligned}
\label{eq:dna_relative_frobenius_error}
\end{equation}
We report the mean and sample standard deviation of this error over the 32 trials. For top-one family retrieval, each sequence returns the other sequence with the largest sampled fidelity; self-matches are excluded. Retrieval is correct when the two sequences share a parent. These metrics distinguish reproducing the complete similarity matrix from retaining enough information to make a correct retrieval decision. Pairwise error maps show the trial mean of
\(\left|F_{\mathrm{QOS}}^{(Q)}(i,\ell)
-F_{\mathrm{exact}}^{(Q)}(i,\ell)\right|\).

We evaluate compilation costs separately on a smaller data set with two parents and four variants per parent: \(N=8\), \(L=256\), \(k=3\), and \(d=64\).
The mutation probability, GC-content parameter (the expected fraction of guanine and cytosine bases), and phase scale remain \(0.05\), \(0.5\), and \(\alpha=8\), respectively. The joint sequence-\(k\)-mer register contains \(Nd=512\) addresses and requires nine qubits. We evaluate \(Q\in\{1,2,4,8\}\) and \(M\in\{2^6,2^7,2^8,2^9,2^{10}\}\), with \(M_0=M/Q\). For each parameter pair, the direct and fused circuits are built from the same independent sample blocks. The direct circuit contains one phase update per sample; the fused circuit combines updates to repeated addresses within each block.  The measured circuit segment contains the \(Q\) empirical oracle applications and the \(Q-1\) Hadamard transformations between them. We exclude the common preparation of \(\ket{+}_d\) and the final Hadamard to isolate the cost of this segment. Therefore, the reported resource counts and compression ratios do not describe the complete fingerprinting circuit. Omitting the final Hadamard does not change the fingerprint fidelities because it preserves inner products. Circuits are transpiled in Qiskit using \texttt{GenericBackendV2}, a line coupling map, and the basis gates \texttt{id}, \texttt{rz}, \texttt{sx}, \texttt{x}, and \texttt{cx}. We use optimization level 1 and \texttt{seed\_transpiler}=7, and report transpiled depth and the number of \texttt{cx} gates.

\section{Results}
\label{sec:results}
We evaluate offline QOS at two levels. First, we determine whether empirical  phase oracles constructed from a limited number of samples preserve the DNA retrieval result. Second, we measure the circuit cost of implementing these oracles directly and after phase fusion.

\begin{figure*}[t]
    \centering
    \includegraphics[width=\textwidth]
    {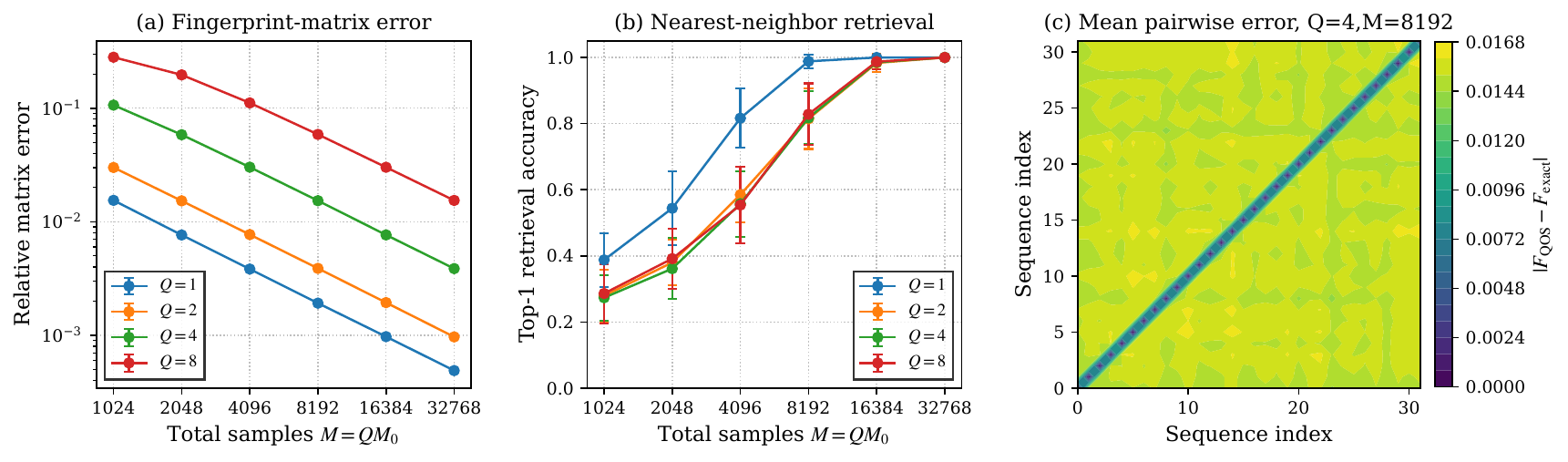}
    \caption{Accuracy of offline QOS for the DNA fingerprinting task.
    Each of the \(Q\) oracle applications is constructed from an independent
    block of \(M_0=M/Q\) samples.
    (a) Mean relative error \(E_F^{(Q)}\) of the fingerprint matrix.
    (b) Mean top-one family-retrieval accuracy. Error bars in (a) and (b)
    show one sample standard deviation over 32 trials.
    (c) Mean pairwise absolute fingerprint error for \(Q=4\) and \(M=8192\).}
    \label{fig:qos_accuracy}
\end{figure*}

Fig.~\ref{fig:qos_accuracy}(a) shows that increasing the sample count
systematically improves the fingerprint approximation. For one oracle
application, the relative error decreases from \(1.54\times10^{-2}\) at
\(M=2^{10}\) to \(4.89\times10^{-4}\) at \(M=2^{15}\). Over this range, the
observed error decreases approximately as \(M^{-1}\).

When the encoding is applied repeatedly, the total sample count is divided among the oracle applications. Each empirical oracle is consequently
constructed from \(M_0=M/Q\) samples, and the sampling errors propagate through the quantum transformations. At \(M=2^{15}\), the errors for \(Q=2,4,8\)
are \(9.68\times10^{-4}\), \(3.87\times10^{-3}\), and \(1.54\times10^{-2}\), respectively. For \(Q\geq2\), the observed scaling is approximately \(E_F^{(Q)}\propto Q^2/M\). This behavior agrees qualitatively with the QOS composition analysis, which predicts that maintaining a fixed accuracy across repeated oracle applications requires a larger total sample count~\cite{Zhao2026QOS}. Here, we evaluate this effect at the application level through the relative error of the resulting fingerprint matrix.

One of the most important results is that retrieval becomes accurate with substantially fewer samples than are required to reproduce the complete fingerprint matrix. The data set contains $32(512-4+1)=16\,288$ \(k\)-mer windows. With one oracle application, \(M=8192\) samples, about half the number of available windows, already give a mean retrieval accuracy of \(0.99\). 

This separation is also visible for repeated encoding. At \(M=2^{13}\), retrieval accuracy is between \(0.81\) and \(0.83\) for \(Q=2,4,8\), even though the individual oracle applications use progressively smaller sample
blocks. At \(M=2^{14}\), all configurations exceed \(0.98\) accuracy. In the case \(Q=8\), this result is obtained with only \(M_0=2048\) samples per oracle application, one eighth of the number of windows in the complete corpus. All configurations reach perfect mean retrieval at \(M=2^{15}\).

The representative error map in Fig.~\ref{fig:qos_accuracy}(c) shows the remaining pairwise errors for \(Q=4\) and \(M=8192\). The mean off-diagonal absolute error is \(1.54\times10^{-2}\). The diagonal error is zero because both the exact and sampled states have unit self-fidelity. Thus, the oracle can retain the family structure needed for retrieval even when individual similarity values remain approximate.

We next evaluate the cost of implementing the sampled oracles. For each \((Q,M)\), the direct and fused circuits use the same independent sample blocks and implement exactly the same sampled computation. The direct construction contains one phase operation per sample. The fused construction combines all
updates to the same address within a block into one accumulated phase. 

We define the \texttt{cx}-gate compression ratio as
$
R_{\mathrm{CX}}
=
\frac{G_{\mathrm{CX,direct}}}
     {G_{\mathrm{CX,fused}}},
$
where \(G_{\mathrm{CX,direct}}\) and \(G_{\mathrm{CX,fused}}\) are the transpiled \texttt{cx} counts before and after fusion.

\begin{figure*}[t]
    \centering
    \includegraphics[width=\textwidth]
    {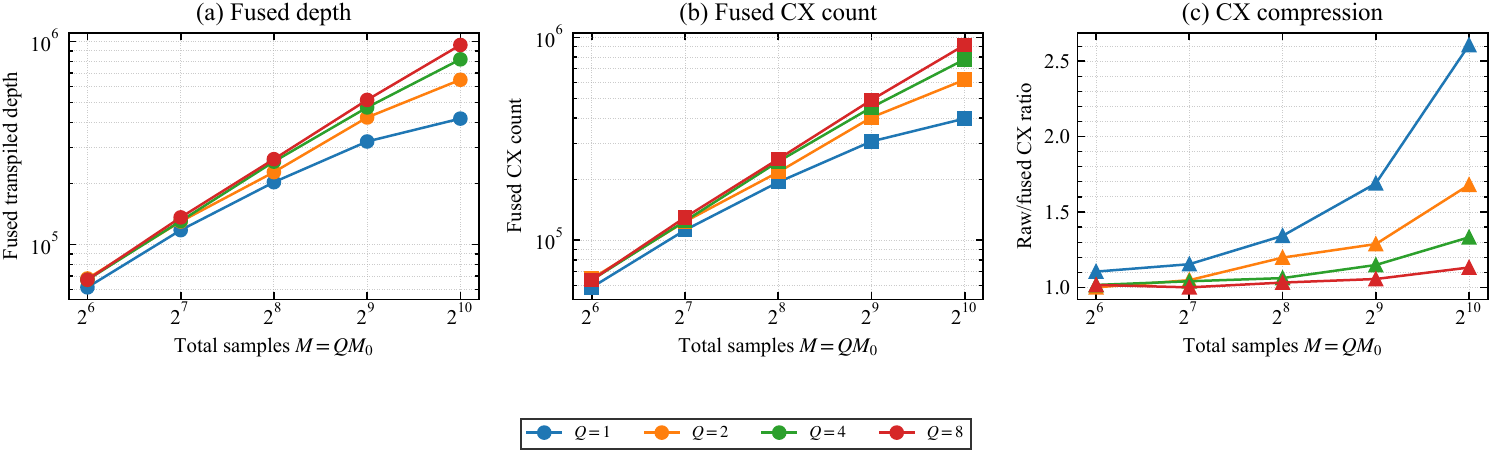}
    \caption{Transpiled resources for the offline QOS fingerprinting circuit.  Direct and fused circuits use the same \(Q\) independent sample blocks and
    the same total sample count \(M=QM_0\). (a) Depth of the fused circuit. (b) Transpiled \texttt{cx} count of the fused circuit. (c) Compression ratio \(R_{\mathrm{CX}}\) relative to the direct sample-by-sample construction.}
    \label{fig:qos_resources}
\end{figure*}

At fixed \(M\), the direct circuit cost is nearly independent of the number of oracle applications because the total number of phase operations remains \(M\). At \(M=2^{10}\), the direct circuits have a depth of approximately \(1.09\times10^6\) and contain about \(1.04\times10^6\) \texttt{cx} gates for every value of \(Q\).

Fig.~\ref{fig:qos_resources} shows that fusion reduces this cost by combining samples that select the same address. For \(Q=1\) and \(M=2^{10}\), the 1024 operations reduce to 393 distinct address phase operations. The fused circuit has depth \(417\,523\) and \(398\,000\) \texttt{cx} gates, giving \(R_{\mathrm{CX}}=2.608\). Fusion becomes less effective when the samples are divided among multiple oracle applications. At \(Q=8\), each block contains only \(M_0=128\) samples, but repeated addresses can be combined only within the same block. The fused composition therefore retains 905 phase operations across the eight blocks. Its depth is \(962\,437\), its \texttt{cx} count is \(917\,249\), and its compression ratio decreases to \(1.132\). The logical phase operation ratios, \(2.606\) for \(Q=1\) and \(1.131\) for \(Q=8\), closely match the transpiled \texttt{cx} ratios.

At fixed \(Q\), increasing the sample count creates more repeated addresses and improves the relative compression obtained by fusion. At the same time, the number of distinct addresses and the absolute circuit cost continue to grow. The resulting circuits remain very large. Even after fusion, the depths at
\(M=2^{10}\) range from \(4.18\times10^5\) to \(9.62\times10^5\). Offline QOS is therefore impractical on current quantum computing devices under this implementation, despite using relatively small sample blocks. \\[0.4em]
\noindent
\textbf{Practical Consideration IX: The application may converge before the oracle does.}
\textit{The DNA retrieval task reaches nearly perfect accuracy while the fingerprint matrix still contains sampling error. In particular, repeated encoding achieves more than \(0.98\) retrieval accuracy with only \(M_0=2048\) samples per oracle application, compared with \(16\,288\) \(k\)-mer windows in the complete corpus. A downstream task with a sufficient decision margin can therefore use a highly approximate oracle constructed from a small sample count. This is an important strength of QOS for large data sets. However, small sample counts do not yet produce small quantum circuits. The fused implementations, studied here, still require depths of several hundred thousand gates and remain impractical on current quantum computing devices.}

\section{Discussion and Conclusion}
\label{sec:conclusion}
QOS was recently proposed as a method for processing massive classical data through sampled quantum oracle access, with strong resource advantages for selected applications in its online computational model~\cite{Zhao2026QOS}. Its main benefit is that an oracle can be approximated from samples of a data distribution without loading the complete data set into a quantum data structure. As demonstrated by the DNA retrieval experiment, the downstream application may become accurate before the empirical oracle has fully converged. This allows useful results to be obtained from sample counts much smaller than the number of observations in the original data set.

Realizing this advantage on current quantum computers remains difficult. Online QOS requires the quantum state to remain coherent throughout sample delivery, classical control, and all quantum gate execution. It also requires an interactive execution model in which incoming samples can control a computation already in progress. Current quantum computing systems are dominated by batched execution and do not generally provide this combination of coherence time and real-time control. Online QOS is therefore not presently practical on these platforms.

We introduced an offline realization that can be expressed within the current quantum circuit execution model. Each oracle application receives an independent sample block, which is summarized on the classical host and compiled into an empirical oracle before execution. This construction preserves the sampled oracle approximation and the structure of the downstream query. However, it does not retain the machine size advantage proved for online QOS and results in deep quantum circuits.

For all these reasons, QOS remains a forward-looking data access model rather than a method that can be effectively deployed on current quantum computing devices. The offline construction provides a way to study its sampling behavior, application accuracy, and compilation costs using current programming tools, but it does not remove the underlying hardware limitations. Making QOS practical will require lower depth synthesis of  oracle updates, compiler transformations that exploit repeated addresses and commuting operations, coherence or fault-tolerance sufficient for the complete streaming computation, and low-latency runtime interfaces that can modify an active quantum program in response to arriving classical samples.

\begin{credits}
\subsubsection{\ackname}
This work was supported by the Wallenberg Centre for Quantum Technology (WACQT) funded by the Knut and Alice Wallenberg Foundation (KAW).
\end{credits}

 \bibliographystyle{splncs04}
 \bibliography{QuantumSketching}

\end{document}